\documentclass[twocolumn,prb,superscriptaddress,showpacs,amsmath,amssymb]{revtex4}
\usepackage{bm,graphicx,dcolumn}
\newcommand{\pdx}{\partial_{x}}

\begin{document}
\title{Theory of vortex structure in Josephson junctions with multiple tunneling
channels: Vortex enlargement as a probe of $\pm s$-wave superconductors}

\affiliation{
CCSE, Japan Atomic Energy Agency, 
6-9-3 Higashi-Ueno Taito-ku, Tokyo 110-0015, Japan}
\affiliation{
Institute for Materials Research, Tohoku University, 2-1-1 Katahira
Aoba-ku, Sendai 980-8577, Japan} 
\affiliation{
CREST(JST), 4-1-8 Honcho, Kawaguchi, Saitama 332-0012, Japan}
\affiliation{
JST, TRIP, 5 Sambancho Chiyoda-ku, Tokyo 102-0075, Japan}
\author{Yukihiro Ota}
\affiliation{
CCSE, Japan Atomic Energy Agency, 
6-9-3 Higashi-Ueno Taito-ku, Tokyo 110-0015, Japan}
\affiliation{
CREST(JST), 4-1-8 Honcho, Kawaguchi, Saitama 332-0012, Japan}
\author{Masahiko Machida}
\affiliation{
CCSE, Japan Atomic Energy Agency, 
6-9-3 Higashi-Ueno Taito-ku, Tokyo 110-0015, Japan}
\affiliation{
CREST(JST), 4-1-8 Honcho, Kawaguchi, Saitama 332-0012, Japan}
\affiliation{
JST, TRIP, 5 Sambancho Chiyoda-ku, Tokyo 102-0075, Japan}
\author{Tomio Koyama}
\affiliation{
Institute for Materials Research, Tohoku University, 
2-1-1 Katahira Aoba-ku, Sendai 980-8577, Japan}
\affiliation{
CREST(JST), 4-1-8 Honcho, Kawaguchi, Saitama 332-0012, Japan}
\author{Hideki Matsumoto}
\affiliation{
Institute for Materials Research, Tohoku University, 
2-1-1 Katahira Aoba-ku, Sendai 980-8577, Japan}
\affiliation{
CREST(JST), 4-1-8 Honcho, Kawaguchi, Saitama 332-0012, Japan}
\date{\today}

\begin{abstract}
We theoretically study Josephson vortex structures in Josephson
 junctions which have multiple tunneling channels caused by multiple
 superconducting gaps. 
 Deriving ``coupled sine-Gordon equations'' from the free energy
 taking account of the multiple tunneling channels, we examine two
 typical situations, a heterotic junction
 composed of multigap superconductor, insulator, and
 single-gap superconductor, and a grain-boundary junction formed by two
 identical multigap superconductors.
 Then, we reveal in both situations that the magnetic field distribution
 of the Josephson vortex 
 for $\pm s$-wave superconductivity is more enlarged than
 that for $s$-wave without sign change between the order parameters.  
 Its mechanism is ascribed to a cancellation of the multiple Josephson currents. 
 We display such an anomalous Josephson vortex and suggest how to evaluate
 the enlargement.
\end{abstract}

\maketitle

The discovery of iron-based high-$T_{\rm c}$
superconductor\,\cite{Kamihara;Hosono:2008,Rotter;Johrendt:2008,Ren;Zhao:2008,Ogino;Shimoyama:2009,Tamegai;Eisaki:2008,Otabe;Ma:2009}
has triggered a numerous number of studies on its superconducting mechanism
and properties. 
It is now well-known through various
experiments\,\cite{Ding;Wang:2008,Kawabata:2008,Hashimoto;Matsuda:2009}
that multiple bands contribute to the 
superconductivity and multiple superconducting full ($s$-wave) gaps open
below the transition temperature. 
On the other hand, several theoretical
works\,\cite{Mazin;Du:2008,Kuroki;Aoki:2008,Bang;Choi:2008,Nagai;Machida:2008,Kuroki;Aoki:2009,Nakai;Machida:2009} have
proposed that a sign change occurs between the $s$-wave gaps 
when a strong repulsion works between the quasiparticles on the
disconnected Fermi surfaces. 
The symmetry with such a sign change has been called
$\pm s$-wave\,\cite{Mazin;Du:2008,Kuroki;Aoki:2008,Bang;Choi:2008,Nagai;Machida:2008,Kuroki;Aoki:2009,Nakai;Machida:2009}, and its peculiar features have been intensively
explored~\cite{Agterberg;Janko:2002,Onari;Tanaka:2009,Linder;Sudbo:2009,Nagai;Hayashi:2009,Linder;Sperstad;Sudbo:2009,Parker;Mazin:2009,Inotani;Ohashi:2009,Golubov;Tanaka:2009,Ota;Matsumoto:2009,Chen:2009,Ota;Koyama:2009}.  

In cuprate high-$T_{\rm c}$ superconductors, the experimental quest for the 
superconducting gap symmetry has a long
history\,\cite{Tsuei;Kirtley:2000}, in which an epoch-making work was
the detection of a half quantized vortex in corner or
tri-crystalline junctions\,\cite{Tsuei;Ketchen:1994}.    
The discovery was so conclusive that such a measurement
has been regarded as the most reliable way to confirm unconventional
pairing symmetry since then.  
Is such a type of phase sensitive measurement also available for
identifying $\pm s$-wave symmetry in iron-based superconductors?
The answer is not so simple\,\cite{Parker;Mazin:2009}, because it seems to be rather
difficult for $s$-wave case to detect the sign change in spatially twisted
geometries. 

In this paper, we propose an alternative way based on the observation of
a Josephson vortex to identiy $\pm s$-wave symmetry. 
The size of the Josephson vortex unexpectedly enlarges for $\pm s$-wave
compared to the size estimated without the sign change. 
Such an enlargement is widely observable in various junction configurations, e.g., 
a heterotic junction composed of an iron-based superconductor, an insulator, and 
a single-gap superconductor (SIS)\,\cite{Ota;Matsumoto:2009}, a grain-boundary 
junction formed by two iron-based superconductor
grains\,\cite{Ota;Koyama:2009}, 
an intrinsic Josephson junction only for highly anisotropic
compounds\,\cite{Ogino;Shimoyama:2009}, and so on.  
The detection will be possible if one uses the scanning
superconducting quantum interference device\,\cite{Kirtley;Wind:1995}. 
In this paper, we derive the ``coupled sine-Gordon equations'' for the
Josephson junctions with multiple tunneling paths stemming from the
multigap character. 
The equations predict an anomalous structure for the Josephson vortex in
the $\pm s$-wave case, in which the sign of the Josephson critical
current density depends on the tunneling channel. 

\begin{figure}[bp]
\centering
(a)\!\!\!\!\!\scalebox{0.187}[0.187]{\includegraphics{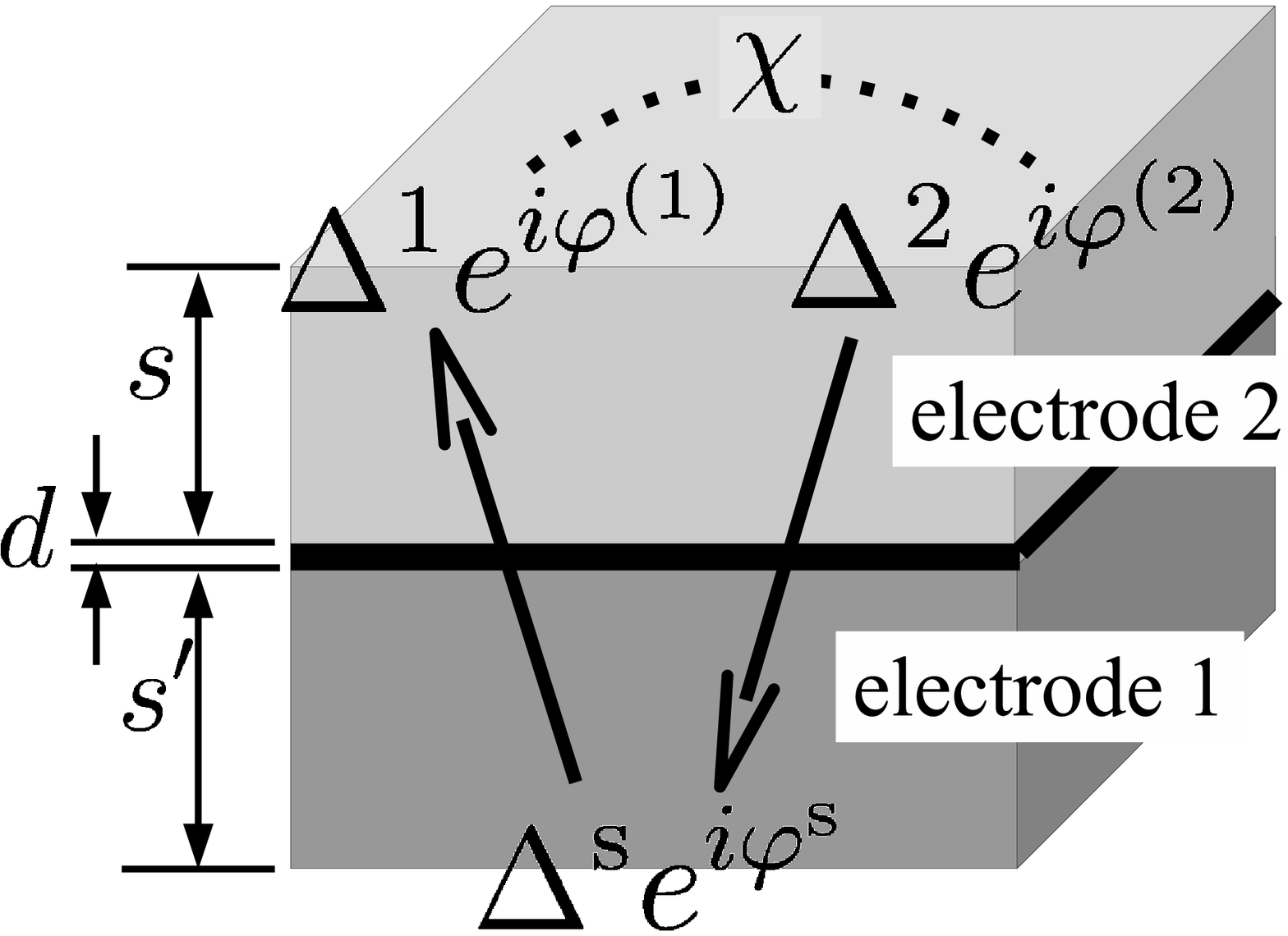}}
(b)\!\!\!\!\!\scalebox{0.187}[0.187]{\includegraphics{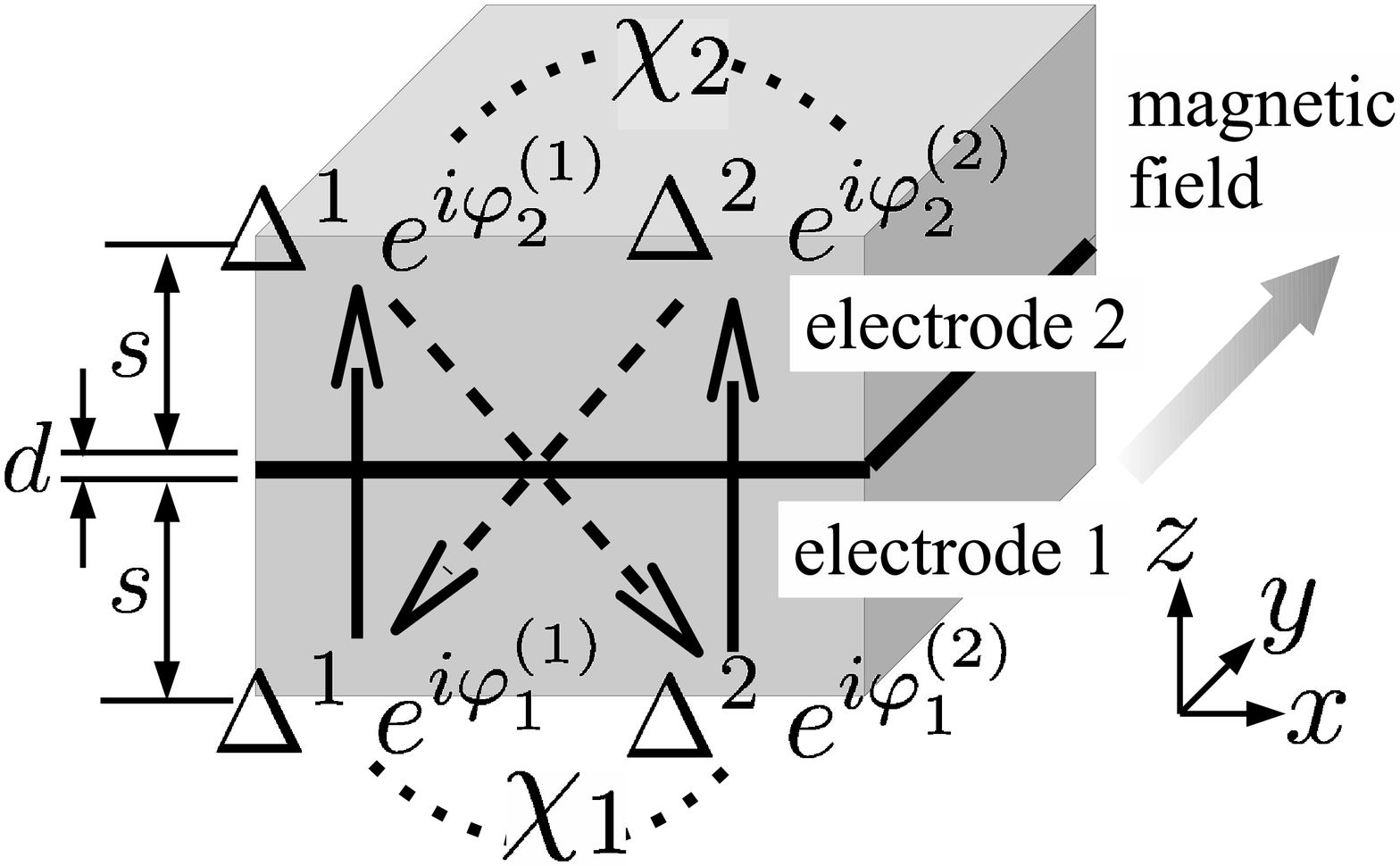}}
\caption{Schematic for Josephson junctions with multiple
 tunneling channels. (a) A heterotic junction between single- and
 two-gap superconductors. (b) A grain-boundary junction between two-gap
 superconductors.} 
\label{fig:schematic_j}
\end{figure} 

The theory of Josephson junctions with multiple tunneling channels is
in great demand for examining and understanding weak link properties of
multi-gap superconductors.  
A theoretical development was done by 
Brinkman {\it et al.}\,\cite{Brinkman;Andersen:2002} and Agterberg 
{\it et al.}\,\cite{Agterberg;Janko:2002}, as for $\mbox{MgB}_{2}$ and
$\mbox{NbSe}_{2}$.  
The modification in the conventional Ambegaokar-Baratoff
relation\,\cite{Ambegaokar;Baratoff:1963} was shown in these
literatures. 
In addition, proximity effects were studied in a heterotic
structure composed of a normal metal and a multuigap
superconductor\,\cite{Brinkman;Kupriyanov:2004}. 
The observation of collective modes in two-gap superconductors via
Josephson junctions was also proposed\,\cite{Anishchanka;Efetov:2007}. 
We note that a peculiar effect of the sign change between the superconducting
gaps on the Josephson current was suggested by Agterberg 
{\it et al.}\,\cite{Agterberg;Janko:2002} in a context irrelevant to
iron-pnictide superconductors.  
After the discovery of iron-pnictide superconductors, the importance of
examining the effects of such a sign change grows significantly. 
A large amount of studies about Josephson junctions or tunneling 
spectroscopy have been reported, e.g., the Andreev bound
states\,\cite{Onari;Tanaka:2009,Linder;Sudbo:2009,Nagai;Hayashi:2009,Golubov;Tanaka:2009}, 
the dc-Josephson effect\,\cite{Linder;Sperstad;Sudbo:2009},
and the Riedel anomaly\,\cite{Inotani;Ohashi:2009}. 
On the other hand, a theoretical research about the magnetic properties of
Josephson junctions with multiple tunneling channels has been never so
far studied, except for our previous work\,\cite{Ota;Matsumoto:2009}. 
Thus, we develop theory of Josephson vortex in such a system on
the basis of a microscopic approach for Josephson
junctions\,\cite{Ota;Matsumoto:2009,Machida;Tachiki:2000}. 

First, we examine the heterotic SIS junction\,\cite{Ota;Matsumoto:2009}.
The situation is shown in
Fig.\,\ref{fig:schematic_j}(a), where the electrode $2$ ($1$) is a
two-(single-)gap superconductor with width $s$ ($s^{\prime}$), 
and the superconducting phases are expressed as $\varphi^{(1)}$ and
$\varphi^{(2)}$ ($\varphi^{\rm s}$). 
The free energy density on the
$zx$ plane\,\cite{Ota;Matsumoto:2009,Machida;Tachiki:2000}
is given by 
\begin{equation}
 \mathcal{F} 
=
\frac{s^{\prime}}{8\pi\lambda^{\prime\,2}}
(a^{x})^{2}
+
\sum_{i=1}^{2}
\frac{s}{8\pi\lambda_{i}^{2}}
(a^{x}_{i})^{2}
+V_{{\rm J}}
+ \frac{d}{8\pi}(B^{y})^{2},
\label{eq:efflag_het} 
\end{equation} 
where
\begin{eqnarray}
V_{{\rm J}}
&=&
-\sum_{i=1}^{2}\frac{\hbar j_{i}}{e^{\ast}}\cos\theta^{(i)}
-\frac{\hbar J_{{\rm in}}}{e^{\ast}}\cos\chi
\label{eq:jenergy_het}, \\
 \theta^{(i)} 
&=& 
\varphi^{(i)} - \varphi^{{\rm s}}  
- \frac{e^{\ast}d}{\hbar c} A^{z}
\label{eq:inv_pd_het}, \\
\chi 
&=&\varphi^{(1)}-\varphi^{(2)}
=\theta^{(1)}-\theta^{(2)}
\label{eq:def_rltp_het}. 
\end{eqnarray} 
The label $i$ represents the band index ($i=1,2$). 
We define $a^{x}$ and $a^{x}_{i}$ as, respectively, 
\(
a^{x}
=
(\hbar c/e^{\ast}) \pdx\varphi^{\rm s} - A_{1}^{x}
\) and 
\(
a^{x}_{i} 
=
(\hbar c/e^{\ast}) \pdx\varphi^{(i)} - A_{2}^{x}
\), where 
\(
e^{\ast}=2e
\). 
The penetration depth of the superconducting state in the electrode $2$
($1$) is written as $\lambda_{i}$ ($\lambda^{\prime}$). 
The third term in Eq.\,(\ref{eq:efflag_het}) describes the Josephson
coupling energy. 
As shown in Eq.\,(\ref{eq:jenergy_het}), the first term represents the contribution
from two different tunneling channels, $j_{1}$ and $j_{2}$, and the
last one is the internal Josephson 
coupling microscopically originating from an inter-band
interaction\,\cite{Ota;Matsumoto:2009,Gurevich;Vinokur:2003}.  
The sign of $J_{\rm in}$ determines the relative phase
difference, $\chi$ between the two order parameters in the two-gap
superconducting electrode $2$.  
When $J_{\rm in}<0$, the preferable value of $\chi$ becomes $\pi$,
which corresponds to $\pm s$-wave.  
The final term in Eq.\,(\ref{eq:efflag_het}) is the magnetic field energy, in which
\(
B^{y} = d^{-1}(A^{x}_{2}-A^{x}_{1})-\pdx A^{z}
\).  

\begin{figure}[tp]
\centering
(a)\!\!\!\scalebox{0.512}[0.512]{\includegraphics{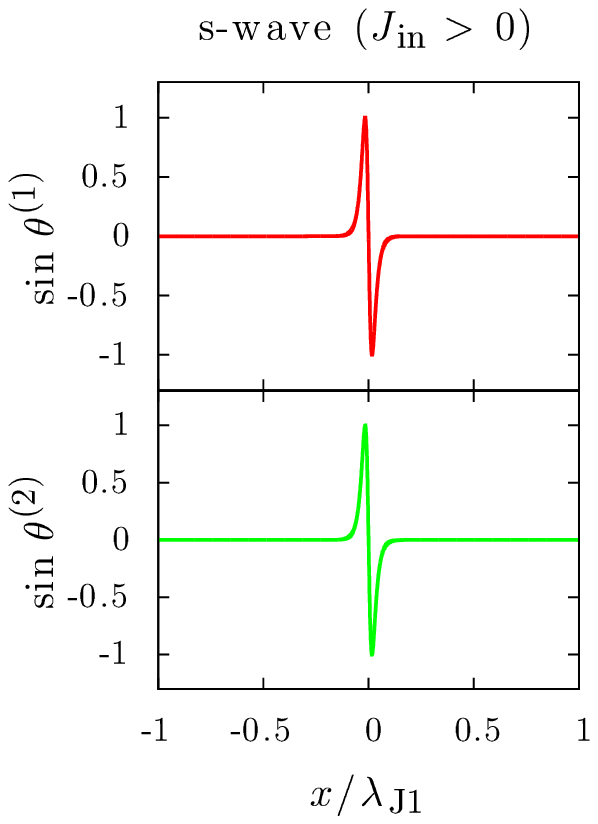}}\,\,\,
(b)\!\!\!\scalebox{0.512}[0.512]{\includegraphics{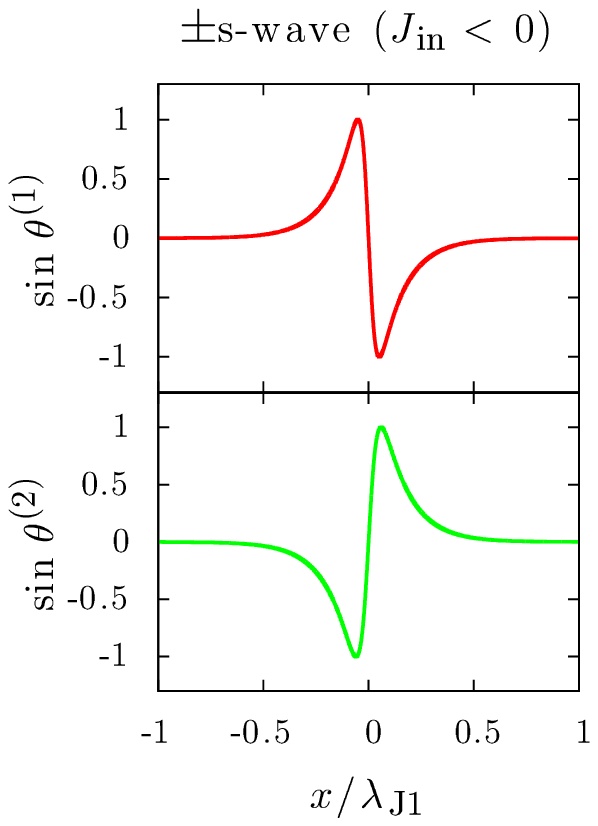}}\\\vspace{4mm}
(c)\!\!\!\scalebox{0.612}[0.612]{\includegraphics{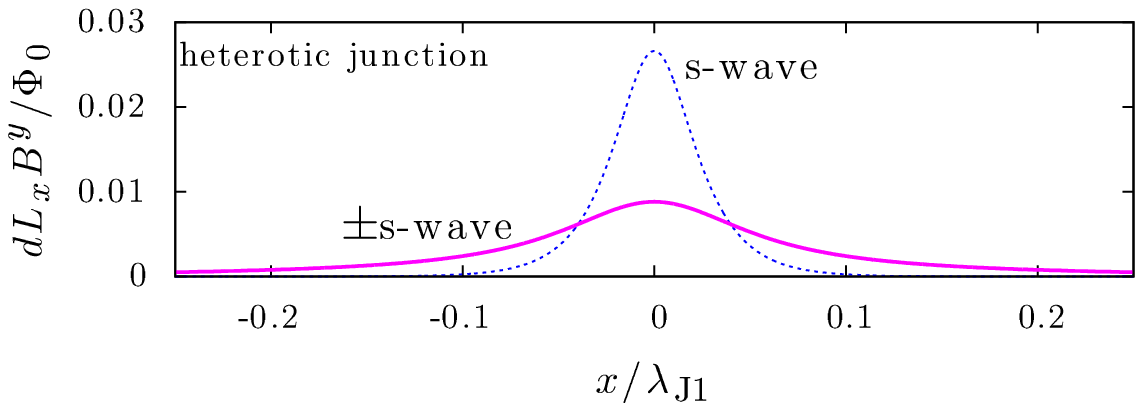}}
\caption{(Color online) The single-vortex solution for the heterotic junction. We set 
$L_{x}=2.5\lambda_{{\rm J}1}$, 
 $\eta^{\prime}=10^{3}$, $\eta_{1}=10^{3}$, $\eta_{2}=1.56\times 10^{3}$, $j_{2}/j_{1}=0.8$, and
 $|J_{\rm in}|/j_{1}=5.0$. (a) $s$-wave ($J_{\rm in}>0$). (b)
 $\pm s$-wave ($J_{\rm in}<0$). (c) The magnetic field penetrating into  
the junction. The solid line is for
 $\pm s$-wave, while the dash one is for $s$-wave.}
\label{fig:het_sv_result}
\end{figure}

Using Eq.\,(\ref{eq:inv_pd_het}) and the Euler-Lagrange equations with respect to
$A^{x}_{\ell}$, we have
\begin{equation}
 \sum_{i=1}^{2}\frac{\bar{\eta}}{\eta_{i}}\pdx\theta^{(i)}
= (1+\eta^{\prime}+\bar{\eta}) \frac{e^{\ast}d}{\hbar c}B^{y},
\label{eq:gJR_het}
\end{equation}
where 
\(
\eta^{\prime} = \lambda^{\prime\,2}/s^{\prime}d
\), 
\(
\eta_{i} = \lambda_{i}^{2}/sd
\), and 
\(
\bar{\eta}^{-1} = \eta_{1}^{-1} + \eta_{2}^{-1}
\). 
Equation (\ref{eq:gJR_het}) is called modified Josephson
relation\,\cite{Machida;Tachiki:2000}. 
Combined Eq.(\ref{eq:gJR_het}) with the Euler-Lagrange equations with
respect to $A^{z}$, $\varphi^{(i)}$, and 
$\varphi^{\rm s}$, we have the coupled sine-Gordon
equations as 
\begin{eqnarray}
 \pdx^{2}\theta^{(1)}
&=& \frac{1+\eta^{\prime}+\eta_{1}}
{\lambda_{{\rm J}1}^{2}}\sin\theta^{(1)}
+\frac{1+\eta^{\prime}}{\lambda_{{\rm J}2}^{2}}\sin\theta^{(2)}
\nonumber \\
&&
+{\rm sgn}(J_{{\rm in}})\frac{\eta_{1}}{\lambda_{{\rm in}}^{2}}
\sin(\theta^{(1)}-\theta^{(2)}), 
\label{eq:csineG_1_het}\\
 \pdx^{2}\theta^{(2)}
&=& \frac{1+\eta^{\prime}}{\lambda_{{\rm J}1}^{2}}\sin\theta^{(1)}
+\frac{1+\eta^{\prime}+\eta_{2}}{\lambda_{{\rm J}2}^{2}}\sin\theta^{(2)}
\nonumber \\
&&
-{\rm sgn}(J_{{\rm in}})\frac{\eta_{2}}{\lambda_{{\rm in}}^{2}}
\sin(\theta^{(1)}-\theta^{(2)}),
\label{eq:csineG_2_het}
\end{eqnarray}
where 
\(
\lambda^{-2}_{{\rm J}i}
= 4\pi de^{\ast}j_{i}/\hbar c^{2}
\) and 
\(
\lambda^{-2}_{\rm in} 
= 4\pi d e^{\ast}|J_{\rm in}|/\hbar c^{2}
\). 
The coefficients of the coupling terms have opposite signs. 
Then, we call the equations ``$\pm$ coupled sine-Gordon equations.''
We note that $\pm$ is not relevant to $\pm s$-wave but common for 
Josephson junctions having multiple tunneling channels. 

Figure \ref{fig:het_sv_result} displays a single Josephson vortex
solution numerically obtained from Eqs.\,(\ref{eq:csineG_1_het}) and
(\ref{eq:csineG_2_het}).  
The spatial scale is normalized by $\lambda_{{\rm J}1}$, and the total
length in the direction of $x$-axis is $L_{x}=2.5\lambda_{{\rm J}1}$. 
The boundary condition is given by 
\(
\theta^{(1)}(-L_{x}/2)=0
\), 
\(
\theta^{(1)}(L_{x}/2)=2\pi
\), 
\(
\theta^{(2)}(-L_{x}/2)=\chi_{0}
\), and 
\(
\theta^{(2)}(-L_{x}/2)=\chi_{0}+2\pi
\). 
The internal phase difference $\chi$ is automatically chosen so that the free
energy becomes minimum. 
We initially choose $-\pi\,(0)$ for $\chi_{0}$ when $J_{\rm in}<0\,(>0)$, and 
solve Eqs.\,(\ref{eq:csineG_1_het}) and (\ref{eq:csineG_2_het}) iteratively.
For the junction parameters, we set $\eta^{\prime}=10^{3}$, $\eta_{1}=10^{3}$,
$\eta_{2}=1.56\times 10^{3}$, $j_{2}/j_{1}=0.8$, and 
 $|J_{\rm in}|/j_{1}=5.0$. 
The width of the current core of the Josephson vortex for $\pm s$-wave as shown 
in Fig.\,\ref{fig:het_sv_result}(b) is much wider than that in
Fig.\,\ref{fig:het_sv_result}(a) for $s$-wave without the sign change. 
Moreover, one finds an antisymmetric current pattern for
$\pm s$-wave\,\cite{Ota;Matsumoto:2009}. 
We can find that $\chi$ is slightly modulated around the vortex center,
although it is almost fixed to be a specific constant ($0$ or $\pi$). 
Using Eq.\,(\ref{eq:gJR_het}), we evaluate the magnetic field distribution
around the Josephson vortex, as is shown in
Fig.\,\ref{fig:het_sv_result}(c). 
We find a significantly enlarged distribution for $\pm s$-wave compared
to $s$-wave without the sign change. 

For further understanding of the above enlargement results, we turn back to 
Eqs.\,(\ref{eq:csineG_1_het}) and (\ref{eq:csineG_2_het}). 
When $\chi$ is rigidly fixed as $0$ or $\pi$, we
have the following equation, which are asymptotically valid except for
the Josephson vortex core (i.e., $|x|\to \infty$): 
\begin{equation}
\pdx^{2}\theta^{(1)}
\sim  \tilde{\eta}_{1}(\chi) \sin\theta^{(1)}, 
\quad
\pdx^{2}\theta^{(2)}
\sim  \tilde{\eta}_{2}(\chi) \sin\theta^{(2)}, 
\label{eq:asympto_sineG}
\end{equation} 
where 
\(
\tilde{\eta}_{1}(\chi) 
= (1+\eta^{\prime}+\eta_{1})/\lambda_{{\rm J}1}^{2}
+ \cos\chi \,(1+\eta^{\prime})/\lambda_{{\rm J}2}^{2}
\) and 
\(
\tilde{\eta}_{2}(\chi) 
= (1+\eta^{\prime})/\lambda_{{\rm J}1}^{2}
+\cos\chi \,(1+\eta^{\prime}+\eta_{2})/\lambda_{{\rm J}2}^{2}
\) ($\chi=0,\,\pi$).  
Here, we emphasize that the characteristic spatial scale 
in Eq.\,(\ref{eq:asympto_sineG}) strongly depends on the type of pairing
symmetry. 
Since 
\(
\tilde{\eta}_{i}(\pi) < \tilde{\eta}_{i}(0)
\), we claim that the solutions for $\pm s$-wave are more widely spread
than that for $s$-wave without the sign change. 
This asymptotic analysis well explains Figs.\,\ref{fig:het_sv_result}(a)
and \ref{fig:het_sv_result}(b). 
In addition, using Eq.\,(\ref{eq:gJR_het}), the magnetic field
distribution inside the junction is asymptotically obtained as 
\begin{equation} 
\frac{L_{x}d}{\Phi_{0}}B^{y}
\sim
A\sum_{i=1}^{2}
\frac{\bar{\eta}}{\eta_{i}}
\frac{2\sqrt{\tilde{\eta}_{i}(\chi)}}
{\cosh[\sqrt{\tilde{\eta}_{i}(\chi)}x]}
\label{eq:approx_B_het}
\end{equation}
where 
\(
A =(1+\eta^{\prime}+\bar{\eta})^{-1}L_{x}/2\pi
\). 
Thus, we clearly find that $\pm s$-wave leads to an enlargement of
the magnetic field distribution due to  
\(
\tilde{\eta}_{i}(\pi) < \tilde{\eta}_{i}(0)
\). 
The origin of such an enlargement is the cancellation between multiple
tunneling channels as shown in Fig.\,\ref{fig:het_sv_result}(b). 
In addition, we point out that the relative phase
difference between the superconducting gaps slightly fluctuates around a fixed
value when $|J_{\rm in}| \gg j_{1},j_{2}$\,\cite{Ota;Matsumoto:2009}. 
The asymptotic forms are valid in this case. 
The above qualitative discussion, Eqs.\,(\ref{eq:asympto_sineG}) and
(\ref{eq:approx_B_het}) does not depend on precise values of the 
junction parameters, as long as the condition is satisfied. 
On the other hand, a quantitative evaluation of the magnetic field
distribution requires the detailed information of the junction
parameters. 
We will discuss the quantitative way to identify the symmetry, i.e.,
$s$-wave or $\pm s$-wave at the end of this paper. 

Second, we examine the grain-boundary junction as schematically shown
in Fig.\,\ref{fig:schematic_j}(b). 
Both the electrodes are assumed to be identical (two-gap) superconductors. 
This type of junction is observed in a weak-link between grains
of a polycrystalline iron-based superconductor\,\cite{Tamegai;Eisaki:2008,Otabe;Ma:2009,Ota;Koyama:2009}.  
Alternatively, the situation is theoretically equivalent to
the intrinsic junctions stacked along the $c$-axis. 
The free energy density is basically similar to
Eq.\,(\ref{eq:efflag_het}), but there are two differences. 
The first term in Eq.\,(\ref{eq:efflag_het}) is substituted with 
\(
-\sum_{i=1}^{2}(s/8\pi \lambda_{i}^{2})(a^{x}_{i,1})^{2}
\), where 
\(
a^{x}_{i,1}=(\hbar c/ e^{\ast})\pdx\varphi^{(i)}_{1} - A^{x}_{1}
\). 
The Josephson coupling energy term is replaced by
\begin{eqnarray}
 V_{\rm J}
&=&
-\sum_{i=1}^{2}
\frac{\hbar j_{i}}{e^{\ast}}\cos\theta^{(i)}_{2,1} 
-\frac{\hbar j_{12}}{e^{\ast}}\cos(\theta^{(2)}_{2,1}-\chi_{1})
\nonumber \\
&&
\quad
-\frac{\hbar j_{21}}{e^{\ast}}\cos(\theta^{(1)}_{2,1}+\chi_{1})
-\sum_{\ell =1}^{2}\frac{\hbar J_{\rm in}}{e^{\ast}}\cos\chi_{\ell},
\label{eq:Jc_gj}
\end{eqnarray}
where
\(
\theta^{(i)}_{2,1}
= \varphi^{(i)}_{2} - \varphi^{(i)}_{1} 
- (e^{\ast}d/\hbar c)A^{z}_{2,1}
\) and 
\(
\chi_{\ell} 
= \varphi^{(1)}_{\ell} - \varphi^{(2)}_{\ell} 
\). 
The first (second) term in Eq.\,(\ref{eq:Jc_gj}) is the
intra-(inter-)band Josephson coupling energy between the two
electrodes\,\cite{Ota;Koyama:2009,Brinkman;Andersen:2002}. 
The inter-band Josephson coupling originate microscopically from 
incoherent (momentum non-conserved) tunneling, which 
is the dominant process at rough boundaries. 

Repeating the same treatment as the previous case, we have the
modified Josephson relation, 
\begin{equation}
  \sum_{i=1}^{2}\frac{\bar{\eta}}{\eta_{i}}\pdx\theta^{(i)}_{2,1}
= (1+2\bar{\eta}) \frac{e^{\ast}d}{\hbar c}B^{y}_{2,1}
\label{eq:gJR_grain}
\end{equation}
and the $\pm$ coupled sine-Gordon equations, 
\begin{eqnarray}
 \pdx^{2}\theta^{(1)}_{2,1}
&=&
\frac{1+2\eta_{1}}{\lambda_{{\rm J}1}^{2}}
\sin\theta^{(1)}_{2,1}
+
\frac{1}{\lambda_{{\rm J}2}^{2}}
\sin\theta^{(2)}_{2,1} 
+f^{\rm inter}_{1}
\nonumber \\
&&
+{\rm sgn}(J_{\rm in})\frac{\eta_{1}}{\lambda_{\rm in}^{2}}
(\sin\chi_{2}-\sin\chi_{1}), 
\label{eq:csineG_1_grain}\\
 \pdx^{2}\theta^{(2)}_{2,1}
&=&
\frac{1}{\lambda_{{\rm J}1}^{2}}
\sin\theta^{(1)}_{2,1}
+
\frac{1+2\eta_{2}}{\lambda_{{\rm J}2}^{2}}
\sin\theta^{(2)}_{2,1} 
+f^{\rm inter}_{2}
\nonumber \\
&&
-{\rm sgn}(J_{\rm in})\frac{\eta_{2}}{\lambda_{\rm in}^{2}}
(\sin\chi_{2}-\sin\chi_{1}), \\
\label{eq:csineG_2_grain} 
\pdx^{2}\chi_{1}
&=&
-\frac{\eta_{1}}{\lambda_{{\rm J}1}^{2}}\sin\theta^{(1)}_{2,1}
+\frac{\eta_{2}}{\lambda_{{\rm J}2}^{2}}\sin\theta^{(2)}_{2,1}
\nonumber \\
&&
-\frac{\eta_{1}}{\lambda_{{\rm J}12}^{2}}
\sin(\theta^{(2)}_{2,1}-\chi_{1})
+\frac{\eta_{2}}{\lambda_{{\rm J}21}^{2}}
\sin(\theta^{(1)}_{2,1}+\chi_{1}) 
\nonumber \\
&&
+
{\rm sgn}(J_{\rm in})\frac{\eta_{1}+\eta_{2}}{\lambda_{\rm in}^{2}}
\sin\chi_{1},
\label{eq:csineG_rp_grain}
\end{eqnarray}
where
\begin{equation}
 f^{\rm inter}_{i}
=
\frac{1+\eta_{i}}{\lambda_{{\rm J}12}^{2}}
\sin(\theta^{(2)}_{2,1}-\chi_{1})
+\frac{1+\eta_{i}}{\lambda_{{\rm J}21}^{2}}
\sin(\theta^{(1)}_{2,1}+\chi_{1}). 
\end{equation}
Another relative phase difference $\chi_{2}$ is determined by the
identity  
\begin{equation}
 \theta^{(1)}_{2,1}-\theta^{(2)}_{2,1} 
= \chi_{2} - \chi_{1}.
\label{eq:diff_rpd_grain}
\end{equation}
Equation (\ref{eq:csineG_rp_grain}) can be regarded as the sine-Gordon
equation with respect to an interband phase
difference\,\cite{Gurevich;Vinokur:2003}. 

\begin{figure}[tbp]
\centering
(a)\!\!\!\scalebox{0.512}[0.512]{\includegraphics{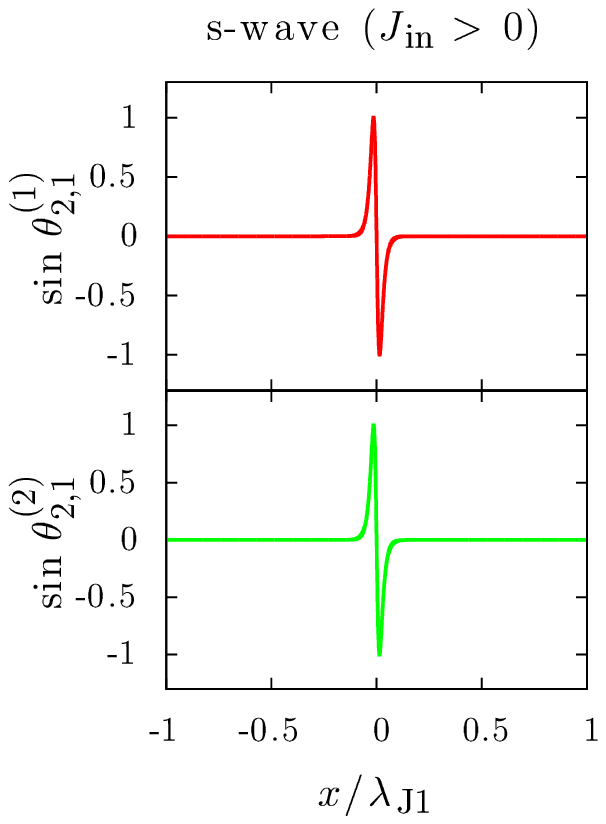}}\,\,\,
(b)\!\!\!\scalebox{0.512}[0.512]{\includegraphics{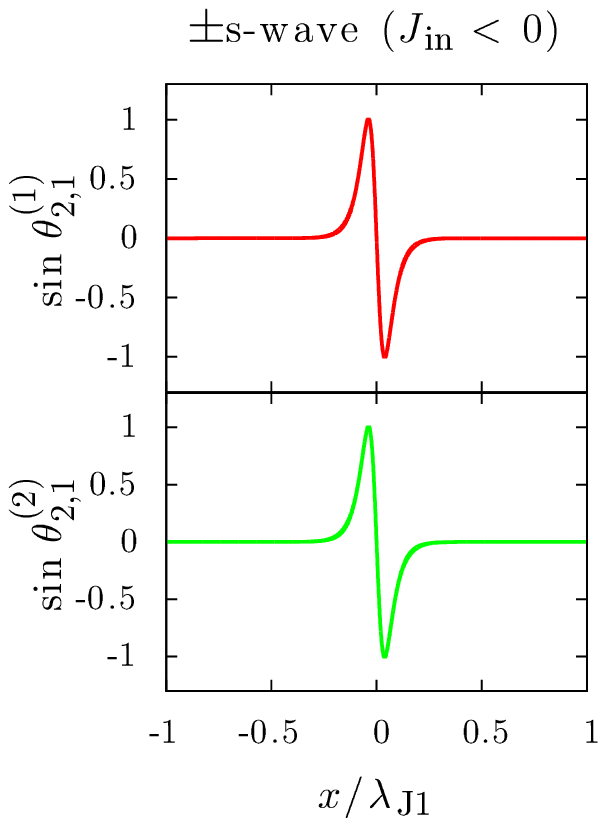}}\\\vspace{4mm}
(c)\!\!\!\scalebox{0.612}[0.612]{\includegraphics{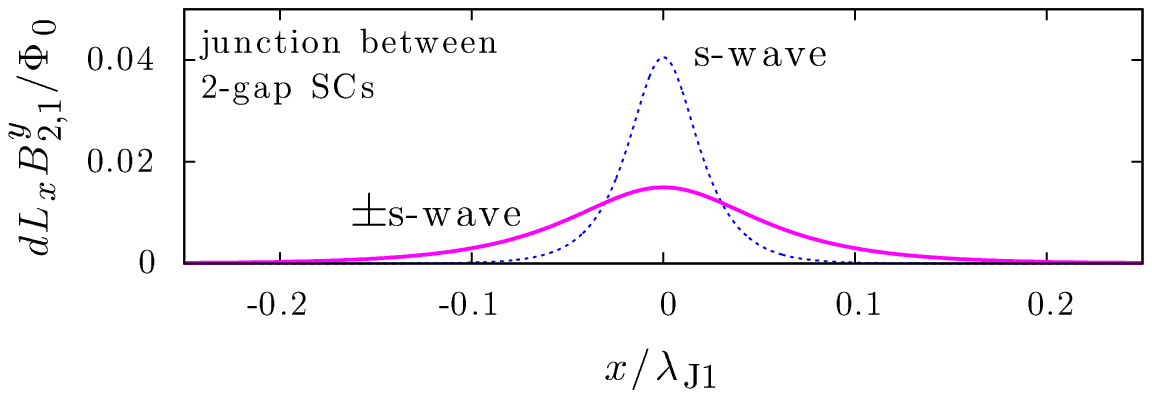}}
\caption{(Color online) The single-vortex solution for the Josephson junction between
 two-gap superconductors. The junction parameters are the same as in
 Fig.\,\ref{fig:het_sv_result}, except for
 $j_{12}/j_{1}=j_{21}/j_{1}=0.6$. 
(a) $s$-wave ($J_{\rm in}>0$). (b)
 $\pm s$-wave ($J_{\rm in}<0$). (c) The
 magnetic field penetrating into the junction. The solid line is for
 $\pm s$-wave, while the dash one is for $s$-wave.}
\label{fig:gj_sv_result} 
\end{figure}

Assuming the physical situation in which superconductivity 
fully grows in each superconducting electrode and $\chi_{1}$ and
$\chi_{2}$ are fixed as the same specific 
value inside the grain, we choose  
\(
\theta^{(i)}_{2,1}(-L_{x}/2)=0
\), 
\(
\theta^{(i)}_{2,1}(L_{x}/2)=2\pi
\), 
and 
\(
\chi_{1}(-L_{x}/2) = \chi_{1}(L_{x}/2) = \chi_{0} 
\) as the boundary condition for the single Josephson vortex. 
We choose $\pi\,(0)$ as $\chi_{0}$ when $J_{\rm in}<0\,(>0)$. 
From Eq.\,(\ref{eq:diff_rpd_grain}), we have
$\chi_{2}(-L_{x}/2)=\chi_{2}(L_{x}/2)=\chi_{0}$. 
Figure \ref{fig:gj_sv_result} displays the single Josephson vortex
solution. 
The ratios, $j_{1}$, $j_{12}/j_{1}=j_{21}/j_{1}=0.6$, and the values of
the other junction parameters are the same as the previous. 
Figures \ref{fig:gj_sv_result}(a) and \ref{fig:gj_sv_result}(b) shows
the shape of the single vortex solution for $s$-wave without the sign
change and $\pm s$-wave, respectively. 
The $\pm s$-wave superconductivity leads to $\chi_{i}=\pi$. 
Thus, Eq.\,(\ref{eq:diff_rpd_grain}) means that no phase difference
between $\theta^{(1)}_{2,1}$ and $\theta^{(2)}_{2,1}$ appears even
though the electrodes are $\pm s$-wave superconductors. 
We find that no anti-symmetric current pattern for $\pm s$-wave appears.  
Figure \ref{fig:gj_sv_result}(c) shows the spatial distribution of the
magnetic field, which is evaluated via Eq.\,(\ref{eq:gJR_grain}). 
We find the enlarging behavior for $\pm s$-wave similar to the hetrotic
SIS case.    
We also find the cancellation between the intra-band and
the inter-band Josephson currents when $J_{\rm in}<0$ (i.e.,
$\chi_{1}=\chi_{2}=\pi$), from Eq.\,(\ref{eq:Jc_gj}). 
Thus, the magnetic field distribution of the Josephson vortex enlarges
in the $\pm s$-wave symmetry. 

\begin{figure}[tbp]
(a)\!\!\!\scalebox{0.52}[0.52]{\includegraphics{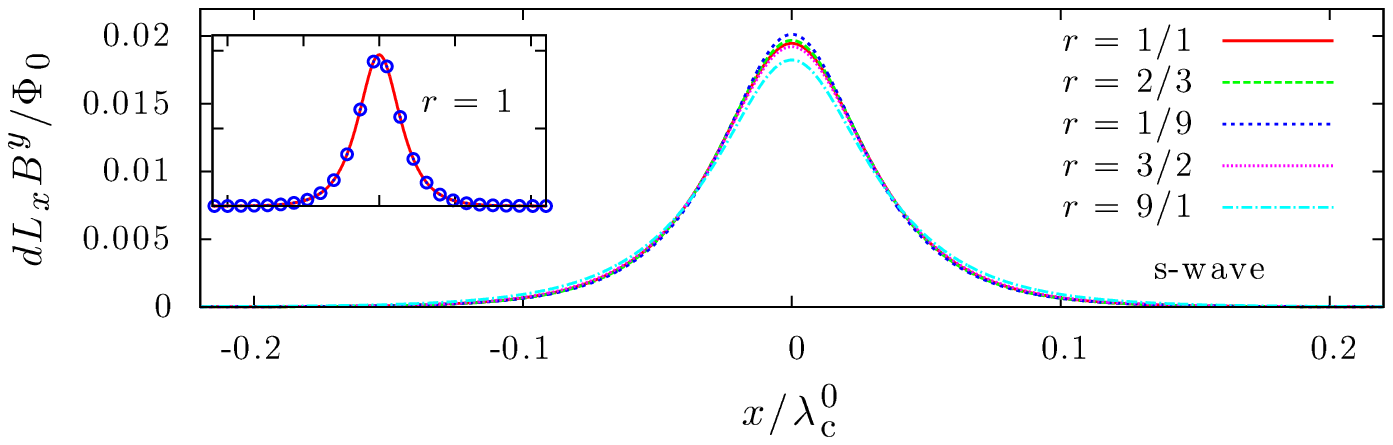}}
(b)\!\!\!\scalebox{0.52}[0.52]{\includegraphics{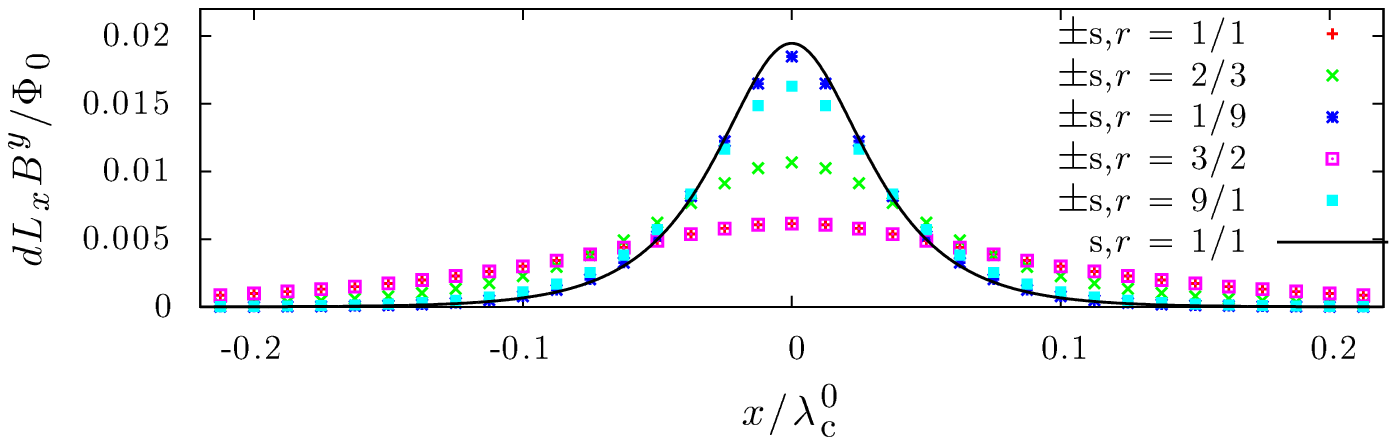}}
\caption{(Color online) The magnetic field penetrating into 
 the heterotic junction, varying $r=r_{{\rm n},1}/r_{{\rm n},2}$. The spatial scale is 
$\lambda_{\rm c}^{0}$, which is evaluated via Eq.\,(\ref{eq:AB_rel}). 
We set $L_{x}=2.5\lambda^{0}_{\rm c}$, $\eta_{1}=10^{3}$, $\eta_{2}=2.0\times 10^{3}$, 
$|J_{\rm in}|/J^{0}_{\rm c}=10$, 
 and $\widetilde{\Delta}^{2}/\widetilde{\Delta}^{1}=0.82$. 
(a) The approximate formula, Eq.\,(\ref{eq:approx_B_het}) for $s$-wave. The inset
 shows the comparison to the numerical solutions for $s$-wave (circle) when $r=1$. 
(b) The comparison of the numerical solutions for $\pm s$-wave to a
 $s$-wave vortex core size. }
\label{fig:appro_mag_het}
\end{figure}

Finally, we discuss how to detect the enlargement of the
Josephson vortex size experimentally.   
Hereafter, we focus on the heterotic junction system.  
A similar discussion is possible for the grain-boundary junction. 
First, we determine a characteristic spatial length for $s$-wave without
the sign change in the case of $j_{1}\simeq j_{2}$. 
The Ambegaokar-Baratoff relation then gives a simple formula, 
\begin{equation}
 R_{\rm n}\frac{\hbar J_{\rm c}^{0}}{e^{\ast}}
\simeq 
\frac{\pi\hbar}{4e^{2}}
\frac{\widetilde{\Delta}^{1}+\widetilde{\Delta}^{2}}{2}. 
\label{eq:AB_rel}
\end{equation}
The quantity $J^{0}_{\rm c}$ is the critical current density in 
the heterotic junction for the the $s$-wave symmetry without sign change.  
The resistance of the junction in the normal state associated with the
tunneling channel $j_{i}$ is written as $r_{{\rm n},i}$. 
We note that 
\(
R^{-1}_{\rm n} = r_{{\rm n},1}^{-1} + r_{{\rm n},2}^{-1}
\) and 
\(
j_{2}/j_{1}
=r_{{\rm n},1}\widetilde{\Delta}^{2}/r_{{\rm n},2}\widetilde{\Delta}^{1}
\). 
We define $\widetilde{\Delta}^{i}$ as 
\(
\widetilde{\Delta}^{i} 
=2\Delta_{\rm S}^{i}K(k_{i})/\pi
\), where 
\(
\Delta_{\rm S}^{i}=\min\{\Delta^{\rm s},\Delta^{i}\}
\), 
\(
\Delta_{\rm L}^{i}=\max\{\Delta^{\rm s},\Delta^{i}\}
\), 
\(
k_{i} = [1-(\Delta_{\rm S}^{i}/\Delta_{\rm L}^{i})^{2}]^{1/2}
\), and $K(k_{i})$ is the complete elliptic integral of the first 
kind\,\cite{Whittaker;Watson:1927}. 
The superconducting gap amplitude in the electrode $1$ ($2$) is written
as $\Delta^{\rm s}$ ($\Delta^{i}$) as shown in
Fig.\,\ref{fig:schematic_j}(a). 
Here, an important point is that the quantity $\widetilde{\Delta}^{i}$
depends on only the superconducting gap amplitudes. 
The direct evaluation of each resistance, 
$r_{{\rm n},1}$ or $r_{{\rm n},2}$, is not practical, but the combined
one $R_{\rm n}$ is measurable in the normal state. 
Thus, one can evaluate $J^{0}_{\rm c}$ from $R_{\rm n}$ and
$\widetilde{\Delta}_{i}$, both of which are supposed to be
experimentally measured, and define a spatial scale as 
\(
\lambda^{0}_{\rm c}
= \sqrt{\hbar c^{2}/4\pi de^{\ast}J^{0}_{\rm c}}
\). 
Normalizing Eqs.\,(\ref{eq:csineG_1_het}) and (\ref{eq:csineG_2_het}) via 
$\lambda^{0}_{\rm c}$, we can find that the equations have a free parameter
$r=r_{{\rm n},1}/r_{{\rm n},2}$.  
Next, we estimate the magnetic field distribution by employing
Eq.\,(\ref{eq:approx_B_het}).  
We also check the dependence of $r$ on the
magnetic field distribution. 
Figure \ref{fig:appro_mag_het}(a) shows distributions of
the magnetic field obtained from Eq.\,(\ref{eq:approx_B_het}) in various $r$ 
for the $s$-wave without the sign change.  
We then find that $r$ dependence of the distribution is
not at all significant.  
Hence, one can adopt the result for $r=1$ as a theoretical prediction
for no sign change. 
Figure \ref{fig:appro_mag_het}(b) presents a comparison with the
$\pm s$-wave case. 
We find that the field distributions in the $\pm s$-wave case are much
wider than that in the case without sign change except for the cases,
e.g., $r=1/9$ or $r=9/1$. 
When $r=1/9$, 
\(
j_{2}/j_{1}\simeq 0.09
\), indicating that one of the multiple tunneling channels
is inactive and the system is approximately described by a
single-channel junction.  
This is not the case of the present iron-based superconductors. 
We emphasize that the magnetic field distribution of the Josephson
vortex for the $\pm s$-wave superconductivity never obeys the
prediction on the basis of the $s$-wave without the sign change except
for such extreme cases. 
In other words, the observed length of the magnetic field extent is much
larger than the theoretical size for the $s$-wave symmetry. 

In conclusion, we studied the single Josephson vortex solutions in the
heterotic SIS Josephson and the grain-boundary junctions, and revealed
an anomalous enlargement of the vortex core size for $\pm s$-wave compared
to the size estimated by the Ambegaokar-Baratoff relation for no sign
change. 
All phenomena were explained on the basis of the cancellation between
different tunneling channels due to the $\pm s$-wave superconductivity. 
As for the heterotic SIS Josephson junction, the cancellation appears 
between the two Josephson currents $j_{1}$ and $j_{2}$. 
On the other hand, the cancellation between the intra- and
inter-grain Josephson currents occurs in the case of the grain-boundary
junction. 
Such a cancellation leads to an effective change in a characteristic
spatial length (e.g., penetration depth). 
Consequently, the Josephson vortex widely provides a reliable way to detect the
gap symmetry in iron-based superconductors. 

The authors (Y.O. and M.M) wish to acknowledge valuable discussion with
S. Shamoto, N. Hayashi, Y. Nagai, H. Nakamura, M. Okumura, N. Nakai, and
R. Igarashi. 
The work was partially supported by Grant-in-Aid for Scientific Research
on Priority Area ``Physics of new quantum phases in superclean
materials'' (Grant No. 20029019) from the Ministry of Education,
Culture, Sports, Science and Technology of Japan.


\begin{thebibliography}{99}
\bibitem{Kamihara;Hosono:2008}
Y. Kamihara, T. Watanabe, M. Hirano, and H. Hosono, 
J. Am. Chem. Soc. {\bf 130}, 3296 (2008). 
\bibitem{Rotter;Johrendt:2008}
M. Rotter, M. Tegel, and D. Johrendt, 
Phys. Rev. Lett. {\bf 101}, 107006 (2008).
\bibitem{Ren;Zhao:2008}
R. Zhi-An, L. Wei, Y. Jie, Yi. Wei, S. Xiao-Li, L. Zheng-Cai,
	C. Guang-Can, D. Xiao-Li, S. Li-Ling, Z. Fang, and Z. Zhong-Xian, 
Chin. Phys. Lett. {\bf 25}, 2215 (2008). 
\bibitem{Ogino;Shimoyama:2009}
H. Ogino, Y. Matsumura, Y. Katsura, K. Ushiyama, S. Horii, K. Kishio,
	and J. Shimoyama, 
Supercond. Sci. Technol. {\bf 22}, 075008 (2009).
\bibitem{Tamegai;Eisaki:2008}
T. Tamegai, Y. Nakajima, Y. Tsuchiya, A. Iyo, K. Miyazawa,
	P. M. Shirage, H. Kito, and H. Eisaki, 
J. Phys. Soc. Jpn. {\bf 77} Supplement C, 54 (2008); 
Physica C {\bf 469}, 915 (2009). 
\bibitem{Otabe;Ma:2009}
E. S. Otabe, M. Kiuchi, S. Kawai, Y. Morita, J. Ge, B. Ni, Z. Gao,
	L. Wang, Y. Qi, X. Zhang, and Y. Ma, 
Physica C {\bf 469}, 1940 (2009). 
\bibitem{Ding;Wang:2008}
H. Ding, P. Richard, K. Nakayama, K. Sugawara, T. Arakane, Y. Sekiba,
	A. Takayama, S. Souma, T. Sato, T. Takahashi, Z. Wang, X. Dai,
	Z. Fang, G. F. Chen, J. L. Luo, and N. L. Wang, 
EPL {\bf 83}, 47001 (2008). 
\bibitem{Kawabata:2008}
A. Kawabata, S. C. Lee, T. Moyoshi, Y. Kobayashi, and M. Sato, 
J. Phys. Soc. Jpn. {\bf 77}, 103704 (2008). 
\bibitem{Hashimoto;Matsuda:2009}
K. Hashimoto, T. Shibauchi, T. Kato, K. Ikada, R. Okazaki, H. Shishido,
	M. Ishikado, H. Kito, A. Iyo, H. Eisaki, S. Shamoto, and
	Y. Matsuda, 
Phys. Rev. Lett. {\bf 102}, 017002 (2009). 
\bibitem{Mazin;Du:2008}
I. I. Mazin, D. J. Singh, M. D. Johannes, and M. H. Du, 
Phys. Rev. Lett. {\bf 101}, 057003 (2008).
\bibitem{Kuroki;Aoki:2008}
K. Kuroki, S. Onari, R. Arita, H. Usui, Y. Tanaka, H. Kontani, and
	H. Aoki, 
Phys. Rev. Lett. {\bf 101}, 087004 (2008); 
{\bf 102}, 109902(E) (2009). 
\bibitem{Bang;Choi:2008}
Y. Bang and H.-Y. Choi, 
Phys. Rev. B {\bf 78}, 134523 (2008).
\bibitem{Nagai;Machida:2008}
Y. Nagai, N. Hayashi, N. Nakai, H. Nakamura, M. Okumura, and M. Machida, 
New. J. Phys. {\bf 10}, 103026 (2008). 
\bibitem{Kuroki;Aoki:2009}
K. Kuroki, H. Usui, S. Onari, R. Arita, and H. Aoki, 
Phys. Rev. B {\bf 79}, 224511 (2009). 
\bibitem{Nakai;Machida:2009}
N. Nakai, H. Nakamura, Y. Ota, Y. Nagai, N. Hayashi, and M. Machida, 
arXiv:0909.1195. 
\bibitem{Agterberg;Janko:2002}
D. F. Agterberg, E. Demler, and B. Janko, 
Phys. Rev. B {\bf 66}, 214507 (2002). 
\bibitem{Onari;Tanaka:2009}
S. Onari and Y. Tanaka, 
Phys. Rev. B {\bf 79}, 174526 (2009).
\bibitem{Linder;Sudbo:2009}
J. Linder and A. Sudb\o, 
Phys. Rev. B {\bf 79}, 020501(R) (2009). 
\bibitem{Nagai;Hayashi:2009}
Y. Nagai and N. Hayashi, 
Phys. Rev. B {\bf 79}, 224508 (2009).
\bibitem{Golubov;Tanaka:2009}
A. A. Golubov, A. Brinkman, Y. Tanaka, I. I. Mazin, and O. V. Dolgov, 
Phys. Rev. Lett. {\bf 103}, 077003 (2009). 
\bibitem{Parker;Mazin:2009}
D. Parker and I. I. Mazin, 
Phys. Rev. Lett. {\bf 102}, 227007 (2009).
\bibitem{Linder;Sperstad;Sudbo:2009}
J. Linder, I. B. Sperstad, and A. Sudb\o, 
Phys. Rev. B {\bf 80}. 020503(R) (2009).
\bibitem{Inotani;Ohashi:2009}
D. Inotani and Y. Ohashi,
Phys. Rev. B {\bf 79}, 224527 (2009). 
\bibitem{Ota;Matsumoto:2009}
Y. Ota, M. Machida, T. Koyama, and H. Matsumoto, 
Phys. Rev. Lett. {\bf 102}, 237003 (2009).
\bibitem{Chen:2009}
C.-T. Chen, C. C. Tsuei, M. B. Ketchen, Z.-A. Ren, and Z. X. Zhao, 
arXiv:0905.3571. 
\bibitem{Ota;Koyama:2009}
Y. Ota, M. Machida, and T. Koyama, 
J. Phys. Soc. Jpn. {\bf 78}, 103701 (2009).  
\bibitem{Tsuei;Kirtley:2000}
C. C. Tsuei and J. R. Kirtley, 
Rev. Mod. Phys. {\bf 72}, 969 (2000).
\bibitem{Tsuei;Ketchen:1994}
C. C. Tsuei, J. R. Kirtley, C. C. Chi, Lock See Yu-Jahnes, A. Gupta,
	T. Shaw, J. Z. Sun, and M. B. Ketchen, 
Phys. Rev. Lett. {\bf 73}, 593 (1994).  
\bibitem{Kirtley;Wind:1995}
J. R. Kirtley, M. B. Ketchen, K. G. Stawiasz, J. Z. Sun,
	W. J. Gallagher, S. H. Blanton, and S. J. Wind, 
Appl. Phys. Lett. {\bf 66}, 1138 (1995).
\bibitem{Brinkman;Andersen:2002}
A. Brinkman, A. A. Golubov, H. Rogalla, O. V. Dolgov, J. Kortus,
	Y. Kong, O. Jepsen, and O. K. Andersen, 
Phys. Rev. B {\bf 65}, 180517(R) (2002). 
\bibitem{Ambegaokar;Baratoff:1963}
V. Ambegaokar and A. Baratoff, 
Phys. Rev. Lett. {\bf 10}, 486 (1963); 
{\bf 11}, 104 (1963). 
\bibitem{Brinkman;Kupriyanov:2004}
A. Brinkman, A. A. Golubov, and M. Yu. Kupriyanov, 
Phys. Rev. B {\bf 69}, 214407 (2004). 
\bibitem{Anishchanka;Efetov:2007}
A. Anishchanka, A. F. Volkov, and K. B. Efetov, 
Phys. Rev. B {\bf 76}, 104504 (2007). 
\bibitem{Machida;Tachiki:2000}
M. Machida, T. Koyama, A. Tanaka, and M. Tachiki, 
Physica C {\bf 331} 85 (2000). 
\bibitem{Gurevich;Vinokur:2003}
A. Gurevich and V. M. Vinokur, 
Phys. Rev. Lett. {\bf 90}, 047004 (2003). 
\bibitem{Whittaker;Watson:1927}
E. T. Whittaker and G. N. Watson, 
{\it A Cource of Modern Analysis} 4th ed. 
(Cambridge University Press, Cambridge, England, 1927), Chap.22. 
\end{thebibliography}
\end{document}